\documentclass[useAMS,usenatbib,usegraphicx]{mn2e}

\usepackage[english]{babel}
\usepackage[utf8x]{inputenc}
\usepackage[T1]{fontenc}
\usepackage{tabularx}
\usepackage{comment}
\usepackage{amsmath}
\usepackage{amssymb}
\usepackage{graphicx}
\usepackage[colorinlistoftodos]{todonotes}
\usepackage[colorlinks=true, allcolors=blue]{hyperref}

\newcommand{\msun}{\ensuremath{\mathrm{M}_\odot}}
\newcommand{\rsun}{\ensuremath{\mathrm{R}_\odot}}

\title{On the formation of neutron stars via accretion-induced collapse in binaries}
\author[Ashley~J.~Ruiter et al.]{A. J. Ruiter$^{1,2,3}$,
L.~Ferrario$^{4}$, K.~Belczynski$^{5}$, I.~R.~Seitenzahl$^{1,2}$, \newauthor
R.~M.~Crocker$^{2}$, A.~I.~Karakas$^{6}$\\
$^{1}$ARC Future Fellow, School of Physical, Environmental and Mathematical Sciences, University of New South Wales,\\ 
Australian Defence Force Academy,
Canberra, ACT 2600, Australia\\
$^{2}$Research School of Astronomy and Astrophysics, Australian National University, Canberra, ACT 0200, Australia\\
$^{3}$ARC Centre of Excellence for All-sky Astrophysics (CAASTRO)\\
$^{4}$Mathematical Sciences Institute, Australian National University, Canberra, ACT 0200, Australia\\
$^{5}$Nicolaus Copernicus Astronomical Center, Polish Academy of Sciences,
           ul. Bartycka 18, 00-716 Warsaw, Poland \\
$^{6}$Monash Centre for Astrophysics, School of Physics and Astronomy, Monash University, VIC 3800, Australia\\
}
\begin{document}
\maketitle

\begin{abstract}
\noindent
We investigate evolutionary pathways leading to neutron star formation through the collapse of oxygen-neon white dwarf (ONe WD) stars in interacting binaries.
We consider (1) non-dynamical mass transfer where an ONe WD approaches the Chandrasekhar mass leading to accretion-induced collapse (AIC) and (2) dynamical timescale merger-induced collapse (MIC) between an ONe WD and another WD. We present rates, delay times, and progenitor properties for two different treatments of common envelope evolution.
We show that AIC neutron stars are formed via many different channels and the most dominant channel depends on the adopted common envelope physics. 
Most AIC and MIC neutron stars are born shortly after star formation, though some have delay times $>10$ Gyr. 
The shortest delay time ($25-50$ Myr) AIC neutron stars have stripped-envelope, compact, helium-burning star donors, though many prompt AIC neutron stars form via wind-accretion from an asymptotic giant branch star. The longest delay time AIC neutron stars, which may be observed as young milli-second pulsars among globular clusters, have a red giant or main sequence donor at the time of NS formation and will eventually evolve into NS + helium WD binaries. We discuss AIC \& MIC binaries as potential gravitational wave sources for LISA. Neutron stars created via AIC undergo a LMXB phase, offering an electromagnetic counterpart for those shortest orbital period sources that LISA could identify.
The formation of neutron stars from interacting WDs in binaries is likely to be a key mechanism for the production of LIGO/Virgo gravitational wave sources (NS-NS and BH-NS mergers) in globular clusters. 
\end{abstract}
\begin{keywords}
binaries: close -- gravitational waves -- stars: evolution -- stars: neutron -- pulsars: general -- white dwarfs
\end{keywords}

\section{Introduction}

It is mostly accepted that accretion from a stellar companion on to a massive carbon-oxygen (CO) white dwarf (WD) either produces a Type Ia supernova, or quiescently burns into an oxygen-neon-rich (ONe)\footnote{We note that a non-negligible fraction of other carbon-burning products, such as Mg-24, are likely to be present in the WD. However, we adopt `ONe WD' to refer to all ONe-rich WDs that also include Mg.} WD, as it approaches the Chandrasekhar mass limit ($\sim 1.4$ \msun).  
The exact burning criteria that determine what actually occurs in Nature is still a subject of debate \citep{nomoto1991a,yoon2005b,schwab2016a}.
It is also not completely clear what happens to even heavier, ONe or hybrid CONe WDs \citep{siess2006a,karakas14b} as they approach the Chandrasekhar mass, 
though the general consensus is that an ONe WD will collapse to form a neutron star (NS) \citep{miyaji1980a}. If the ONe WD is able to achieve central densities of the order of ${\sim} 10^{10}$ g cm$^{-3}$, either through stable accretion from a binary companion \citep{hurley2010a} or through mergers \citep{saio1985a}, electron capture reactions on Ne-20 and Mg-24 likely cause the WD to collapse before a thermonuclear runaway is able to run its full course \citep[see also][]{jones2016a}. 
Thus, in addition to NSs born from traditional iron core-collapse supernovae from massive stars, and electron-capture supernovae born from degenerate ONe cores with zero-age main sequence (ZAMS; for single stars) masses ${\sim} 7-10$ \msun, we also expect a population of NSs born from the accretion-induced collapse of heavy WD stars in interacting binaries. 

NSs created via single  stellar evolution are born shortly after the onset of star formation with a narrow range of delay times, due to the intrinsically massive nature of their progenitors. 
Having said that, there is plenty of evidence showing that core-collapse supernovae from massive stars may commonly occur in binary star systems \citep{smith2014a}, and binary interactions indeed have the effect of pushing the delay time distribution toward larger delay times \citep{zapartas2017a}.  
Binary evolution effects widen the mass range where ONe WDs are expected to form and
enable NSs to be born with a wider range of natal kick velocities \citep[][]{podsiadlowski2004a} and as we will show, delay times. 

The relative fraction of NSs formed via WD accretion in binaries relative to core-collapse NSs formed from massive stars will likely be higher in environments harbouring high stellar densities \citep{ivanova2008a}. 
It is generally assumed that millisecond pulsars (MSPs) are old NSs that were formed in core-collapse events of massive stars and then subsequently spun up via mass accretion \citep[but see also][]{vandenheuvel1984a}. This `recycling' scenario was first suggested by \citet{Backus1982}. However, this is in tension with the evidence that at least some pulsars found among old stellar populations (e.g. globular clusters) exhibit properties consistent with those of young NSs. This anomaly can easily be explained by NSs formed via AIC \citep{lyne1996a,boyles2011a,tauris2013a}. 
\citet{hurley2010a} investigated formation rates of AIC NSs using population synthesis methods \citep[see also][]{ivanova2004a,ferrario2007a,belczynski2010a,liu2018a} and found that AIC systems provide a complementary, if not more important, formation pathway to MSPs than recycling. For a discussion on various formation channels leading to MSPs, we refer the reader to \citet{smedley2017a}. 

Given that current and future deep synoptic sky surveys (e.g. SkyMapper, ZTF, LSST) will continue to unveil an unprecedented number of stellar transients with unknown origin, it is of course worthwhile to predict the most promising observable signatures of different transient candidates; accreting, ultra-massive WDs being the main focus of this work. We distinguish between NSs formed by stable Roche-lobe overflow (RLOF) accretion (accretion-induced collapse, or AIC) and those formed in mergers (merger-induced collapse, or MIC) for the sake of predictions: the former may potentially be observed during their last mass transfer phase shortly before the WD collapses to a NS or after the NS is born during a Low-Mass X-ray Binary (LMXB) phase, while the latter progenitors will be difficult to detect without future, sensitive, space-based gravitational wave observatories such as LISA (see section 3.3).\footnote{Though for MIC progenitors, the merger itself may indeed involve an electromagnetic counterpart, especially if a large amount of mass (on the order of 0.2 \msun) is ejected during the merger \citep[e.g.][]{brooks2017a}.}

From the standpoint of nucleosynthesis, 
demarcating the boundaries separating the conditions for WD collapse versus thermonuclear explosion is critical for delineating the progenitor parameter space between NSs formed in binaries and Type Ia supernovae \citep{saio1985a,maoz2014a}. This has important consequences for chemical evolution and by extension, Galactic archaeology. 
Calculations by \citet{kromer2015a} have shown that lightcurves and spectra of near-Chandrasekhar mass hybrid CONe WDs that undergo a so-called `failed deflagration' could explain faint Type Iax supernovae, such as 2008ha, but do not appear to be able to explain `normal' Type Ia supernovae. 
\citet{marquardt2015a} calculated lightcurves and spectra of thermonuclear supernovae arising from exploding ONe WDs, and though there are some similarities with 1991T-like (very luminous) events, there is no clear match to a particular transient subclass, substantiating the widely-held belief that ONe WDs do not explode \citep[see also][]{wu2017a}. However, A recent multidimensional hydrodynamical study by \citet{jones2016a} found that collapse of an ONe WD to a NS can be avoided in cases where the assumed semi-convective mixing during the electron-capture phase preceding the deflagration is inefficient (resulting in lower central densities at ignition). In these cases the oxygen deflagration results in a weak explosion that leaves behind a bound remnant containing some iron-group material \citep[see also][]{vennes2017a}. The higher central density cases, reflecting very efficient semi-convective mixing, on the other hand showed clear signs of collapse of the WD to a NS.

 An AIC to a NS is indeed expected to be a faint event, certainly compared to `normal' thermonuclear supernovae, and also compared to iron core-collapse supernovae. 
Electron-capture supernova explosions are predicted to have much lower explosion energies \citep[$\lesssim 10^{50}$ erg,][]{dessart2006a,metzger2009a} compared to other core-collapse SNe whose typical explosion energies are (0.6 - 1.2) $\times 10^{51}$ erg. 
In the case of an AIC to a NS, ejecta masses have previously been estimated to be in the range of a few\,$\times 10^{-3}$ to  ${\sim} 0.05$ \msun\ \citep[][]{dessart2006a,darbha2010a,fryer1999a}. 
Even though double NS mergers remain the favoured site for explaining most r-process material \citep{cote2017a}, if the conditions for AIC are similar to those of simulated electron capture SNe, it is reasonable to suppose that AIC events could provide one of the elusive formation sites for the weak r-process \citep{wanajo2011a,cescutti2014a}. 
By providing physical properties of the binary systems that we predict will produce NSs via induced-collapse (both stably-accreting and mergers) from binary evolution population synthesis, we hope to motivate further, detailed studies of interacting  ultra-massive WDs. 
In addition, recent works have proposed that AIC and/or MIC NSs could potentially give rise to other exciting phenomena as well, including fast radio bursts \citep[FRBs,][]{moriya2016a}, magnetars \citep{piro2016a}, and gamma ray bursts \citep{lyutikov2017a}, and they could make an important contribution to gravitational wave astronomy \citealt[see Sec~\ref{sec:GW}, and][]{lipunov2017a}. More specifically: NSs that are formed by induced-collapse channels can contribute toward producing double NS mergers in globular clusters \citep{belczynski2018a}. 

We use the {\sc StarTrack} binary evolution population synthesis code \citep{belczynski2008a,belczynski2002a,ruiter2014a} to determine the birthrates, ages (delay times), and evolutionary pathways of binary star systems that lead to the birth of a NS. We assume two cases for NS formation: 1) an ONe-rich WD in a binary that accretes matter from its companion until  it approaches the Chandrasekhar mass limit and collapses into a NS (accretion-induced collapse, or AIC) and 2) a dynamical merger of two WDs, where at least one of them is already ONe-rich, leading to the birth of a NS via mass accretion (up to Chandrasekhar) onto the more massive WD (merger-induced collapse, or MIC, \citealt{ivanova2008a}). It remains to be established whether the scenarios proposed above lead in reality to a NS rather than e.g. a black hole. For example, a binary system may experience two AIC events: first a WD collapsing to a NS, then the NS collapsing to a BH \citep{belczynski2004a}. In general, the outcome clearly depends on the final mass of the object produced via collapse and on the maximum mass that can be attained by a NS \citep{timmes1996b}. However, this limiting mass is difficult to ascertain because the initial to final mass relationship for NSs is not well constrained by observations \citep[see][and references therein]{margalit2017a}.
In this paper we make the assumption that all such systems will yield NSs, even in the case where the combined mass at time of WD merger exceeds $2.5$ \msun\, which is the canonical upper NS mass limit assumed in {\sc StarTrack} (see Sec\,\ref{sec:results}).
Since the common envelope (CE) phase is the most important source of uncertainty in the formation of WDs in binaries \citep{demarco2009a}, we employ two different parametrizations for CE evolution to bracket the (unknown) uncertainties in the formation of these systems. The two CE treatments are described in Sec\,\ref{sec:bps}.

\section{Model}

\label{sec:bps}

We use the {\sc StarTrack} rapid binary evolution population synthesis code to evolve two populations of binary stars consisting of $12.8 \times 10^{6}$  systems each, starting from the ZAMS.
The code has undergone many updates in recent years. Updates especially relevant for the study of intermediate-mass stars include the addition of a new treatment for CE evolution \citep{xu2010a,dominik2012a}, an updated prescription for the treatment of helium accretion on to WDs \citep{ruiter2014a,kwiatkowski2015a}, and the fact that we now allow a WD to gain mass from a stellar companion that is experiencing wind mass loss at a high rate \citep{belczynski2016b}. 

{\em Initial orbital parameters.} Initial ZAMS star masses are drawn from a 3-component power-law initial mass function based on \citet{kroupa1993a} with $\alpha_{1} = -1.3$, $\alpha_{2} = -2.2$, $\alpha_{3} = -2.35$. The initially more massive star ($M1$) is chosen in the mass range $0.8 - 100$ \msun while the companion's ($M2$)  mass range is $0.5 - 100$ \msun.  $M1$ is drawn directly from the probability distribution function given by our chosen IMF, and $M2$ is calculated by picking a value for the mass ratio $q = 0 - 1$, e.g.\, $M2=qM1$, where $q$ is chosen with a flat probability distribution. We assume circular orbits from the ZAMS rather than the thermal distribution usually adopted in our previous work. This guarantees a higher degree of reproducibility (for comparison purposes) with other population synthesis code data (the majority of binary population synthesis codes assume circular orbits during mass transfer phases, and a circular orbit sometimes must be imposed by hand for a fraction of binaries that would not have circularised naturally due to tidal interactions). 
We note that assuming $e=0$ for every ZAMS binary may not be the most realistic assumption particularly for binaries with large orbital periods \citep[$> 2$ days, see][]{moe2017a}. For interacting binaries, where separations are already rather small, we do not expect a major impact on our final results.  However, we note an overall smaller birthrate for MIC systems when compared to simulations where non-zero initial eccentricity is adopted (see Sec. 4.1).

For the orbital separations on the ZAMS we adopt the canonical initial distribution (flat in the logarithm), up to $10^{5}$ \rsun, with the lower limit set by:
\begin{equation}
a_{\rm min} = f_{\rm amin}\frac{(R_{\rm 1,0} + R_{\rm 2,0})}{(1.0 - e_{0})}
\end{equation}
where $ f_{\rm amin}=2$ is a stellar radius multiplication factor defining the minimum orbital size, $R_{\rm 1,0}$ and $R_{\rm 2,0}$ are the ZAMS radii of stars M1 and M2, respectively, and the initial eccentricity $e_{0}$ is set equal to zero in this paper. 
Both populations are evolved with near-solar (Z=0.02) metallicity. While we do not conduct a parameter study of metallicity here, we note that event rates are generally higher for lower metallicity, for reasons related to wind-mass loss rates, described in \citet{cote2018a}. We discuss the potential impact that initial helium fraction could have in our study in broad terms in section 4.2.  

{\em Common envelope evolution.} A CE is encountered when mass from the donor star is transferred to the accretor on a timescale that is  far too short for the accretor to adjust thermally to the incoming material. As a consequence, this material heats up and swells, filling the Roche lobe. Further mass loss from the donor will form an envelope engulfing both stars. The difference between the two population models is due to the difference in how the binding energy parameter $\lambda$ (see equation 2) is estimated during CE evolution. Though significant progress has been made in quantifying the CE phase numerically over recent years \citep{demarco2011a,ricker2012a,ivanova2013a,ohlmann2016a}, we are far from a comprehensive understanding of this critical phase of binary star evolution. 

In population synthesis studies the CE phase cannot be explicitly calculated but must be somehow parametrized. A common way to do this is to equate the binding energy of the envelope of the mass-losing star 
with the orbital energy of the binary system.
The envelope will then be expelled from the system at the expense of the binary's orbital energy \citep{webbink1984a}, which causes the orbital size to decrease, often drastically. It is reasonable to assume that the binary's post-CE separation is determined by the energy reservoirs in the system that are available, however, we do not know how orbital energy can be transferred to the removal of the envelope; in fact it may not even be a correct assumption \citep{nelemans2005a}. Further, it is not clear which energy reservoirs are even at our disposal \citep[internal energy, ionization energy, enthalpy, see e.g.][]{ivanova2011a}. For this reason the CE efficiency parameter $\alpha_{\rm ce}$ \citep{Livio1988} and the binding energy parameter $\lambda$ \citep{dekool1990}  were introduced. These parameters contain our limited knowledge  on the physics of CE evolution. 

Setting the change in orbital energy  equal to the binding energy gives: 
\begin{equation}
  \alpha_{\rm CE}  ( \frac{G M_{\rm com}  M_{\rm core}}{a_{\rm fin}} - \frac{G M_{\rm com}  M_{\rm core+env}}{a_{\rm init}}  ) =  \frac{G M_{\rm core+env} M_{\rm env} } {\lambda R_{\rm core+env}}, 
  \end{equation} where $M_{\rm com}$ is the mass of the companion star (which is not losing its envelope), $M_{\rm core}$ is the mass of the core of the envelope-losing star, $M_{\rm core+env}$ and $R_{\rm core+env}$ are the total stellar mass and radius of the envelope-losing star, and $a_{\rm fin}$ and $a_{\rm init}$ represent the final and initial orbital separations (post- and pre-CE), respectively. 
 
For Model 1 we adopt the CE formalism employing energy balance \citep{webbink1984a} that historically was the favoured prescription used in binary population synthesis codes with $\alpha_{\rm ce} \lambda = 1$. We refer to Model 1 as the ``classical CE'' model. 
Higher values of $\alpha_{\rm ce} \lambda$ correspond to larger envelope ejection efficiency which, in turn, leads to wider post-CE orbital separations. However, more plausibly, the mass-losing stars' state of evolution will play a role in determining the binding energy of the envelope, and thusly will directly affect the value of the binding energy term. For this reason, we adopt a second CE model that takes the evolutionary stage of the donor into account, while keeping the efficiency parameter, $\alpha_{\rm ce}$, constant (=1). 

It has been clear for many years that adopting a constant value for the binding energy parameter during the CE phase is not the most physical, and therefore probably not the best, assumption \citep{dewi2000a}. It is known however that the parameter $\lambda$ is very sensitive to the structure of the mass-losing star, namely the definition of the core-envelope boundary \citep{dominik2012a}, which is not trivial to prescribe. Nonetheless, we can approximate a value for $\lambda$ based on the evolutionary stage of the donor, envelope mass and radius as calculated from detailed stellar evolution models \citep{xu2010a} in an attempt to ensure that these quantities have an influence on the efficiency with which the envelope is ejected (and thus the post-CE separation). For Model 2 we still use the same energy balance formula (equation 2) but use a more sophisticated approach to estimate the donor binding energy parameter $\lambda$.  We denote this as the ``new CE'' model. Within the context of the energy balance formalism, numerical simulations of common envelope ejection in recent years has revealed that seemingly unphysically high values for ejection efficiency are required to expel the envelope during the CE phase. One possible mechanism that was proposed to help the ejection process is stationary mass outflows \citep{ivanova2011a}. Including such a mechanism, effectively accounting for the enthalpy of the envelope, in the standard energy balance formalism for CE ejection introduces a correction factor by increasing the binding energy term (effectively reducing the binding energy). Such a result is desirable since it enables better agreement with binary population synthesis models and their predicted numbers of black hole LMXBs. 

To estimate the binding energy we use the revised fits of \citet{dominik2012a} based on the data from \citet{xu2010a,xu2010E}. 
In {\sc StarTrack} we have introduced \citep{wiktorowicz2014a} the 
possibility to take into account enthalpy arguments presented by \citet{ivanova2011a}. 
These models allow for a  physical estimate of donor star binding energy that depends on stellar mass, as well as its evolutionary stage and chemical composition (e.g., larger values of $\lambda$ for more evolved donors). In the new CE model, the binding energy parameter value ultimately depends on the evolutionary state of the donor when the CE phase begins.  
Using Table 1 of \citet{ivanova2011a} for intermediate-mass stars as a guide, we increase the binding energy parameter (estimated from our new mass and evolutionary stage--dependent $\lambda$ approach) by a factor of 2, thereby decreasing the binding energy by a factor of 2. 

{\em Stable RLOF on WDs.} For binaries in RLOF (the AIC progenitors), we assume that mass accretion proceeds with the same prescription for ONe WDs as it does for CO WDs \citep[see][for hydrogen-rich and helium-rich accretion, respectively]{ruiter2009a,ruiter2014a}, being dependent on the rate at which mass is transferred toward the accretor, and on the mass of the accreting WD. Thus our adopted prescription is slightly different from the one adopted in \citet{hurley2010a}, which assumes no mass-dependence on the WD accretor for the accretion rate. We also assume that mass transfer of CO (or ONe) on a CO or ONe WD is fully conservative \citep[see][for further details on how mass transfer phases are calculated]{belczynski2008a}. We assume that any CO WD that approaches the Chandrasekhar mass will produce a SN~Ia, regardless of the accretion rate. Thus only ONe WDs are considered to undergo AIC in our simulations (but see Section 3.2).

{\em Stellar winds.} 
{\sc StarTrack} employs various prescriptions from the literature for the treatment of stellar winds. 
For low-mass stars, we adopt the wind prescription of \citet{hurley2002a} \citep[see][for further details]{belczynski2010b}, while for Wolf-Rayet helium stars the wind prescription is based on \citet{hamann1998}, with a metallicity-dependence adopted from \citet{vink2005}. For low- and intermediate-mass evolved stars, it is reasonable to assume that some material lost by the donor may be accreted by a close companion. In fact, such assumptions are likely necessary for re-producing observed properties of carbon-enhanced metal-poor stars \citep{abate2013a}. We assume that a WD can accrete mass from the wind of a nearby companion AGB star donor assuming a Bondi-Hoyle accretion configuration, though this assumption may be conservative, since some simulations have shown wind accretion efficiency from evolved donor stars could be notably higher \citep{mohamed2007a}. For our AIC progenitors with AGB donors, which will lose ${\sim}$a few \msun\ on the AGB, the time-averaged accretion rate is ${\sim}$ $10^{-8}$ \msun\ per year. This AGB wind accretion model is capable of producing AIC NSs from ONe WDs that are born with masses already very close to the Chandrasekhar mass limit. 

{\em Neutron star formation.} There are different ways in which a NS can form from an intermediate-mass star. 
Electron-capture supernovae (ECSNe) form NSs via the collapse of a degenerate ONe core \citep{nomoto1987a,jones2013a,woosley2015a}. The ZAMS mass of an ECSN progenitor is not extremely well-constrained since it will depend on stellar wind mass loss suffered by the star as well as binary interactions, if applicable. However, the helium core mass at the base of the AGB is thought to be $\lesssim 2.5$ \msun\ \citep[see][and references therein]{chruslinska2018a}. 
ECSNe have been predicted to comprise roughly 4 per cent  of all core-collapse supernovae \citep{wanajo2011a}. 

The main difference between the progenitors of ECSNe and AIC is that the ONe core of an ECSN precursor is still surrounded by He- and H-burning shells, which allows the ONe core to grow in mass, collapse, and subsequently suffer a deflagration \citep{hillebrandt1984a}. This different configuration may affect the resulting nucleosynthesis during and possibly also after the formation of the NS. 
In this work we do not discuss general ECSN formation channels but focus on the sub-population of NSs formed through evolutionary channels that involve the collapse of a WD star in an interacting binary. Below, we summarise the conditions for AIC and MIC. 

Today, one of the most promising progenitor scenarios for the formation of Type Ia supernovae is the merger of two CO-rich WDs \citep{nomoto1985a,pakmor2010a,pakmor2012a,ruiter2013a,tanikawa2015a}, with some support for  CO WDs merging with  He-rich WDs \citep{dan2015a,fenn2016a,crocker2017a,brown2017a}.  It is still under investigation whether the WD merger proceeds via `accretion up to Chandrasekhar', in which case the explosion mechanism would likely be similar to the classic single degenerate scenario \citep[e.g.][]{roepke2007c,seitenzahl2013a}, or occur via prompt detonation such as the `helium-ignited violent merger model' of \citet{pakmor2013a} \citep[see also][]{shen2014a}. 
Analogous to CO WDs, in our binary evolution population synthesis simulations concerning ONe WDs we do not differentiate between the two types of merger scenarios, and we include all heavy WD mergers as potential MIC candidates. For the MIC scenario, we assume that a NS is formed any time an ONe-rich WD merges with another WD regardless of its composition or total mass (which always exceeds $1.5$ \msun, see Sec. \ref{sec:results}). 
In our models, ONe WDs only merge with CO WDs or ONe WDs. This is because for other, less-massive WD types (e.g. He-rich), the mass ratio is far from unity and thus when the less-massive WD fills its Roche-lobe mass transfer is dynamically stable, and a merger is avoided in our models \citep[cf.][section 6.2]{cote2018a}. For instance, most of the observed AM\,CVn systems, which are short period binaries where a  He-rich star (likely a white dwarf) transfers mass to its more massive WD companion \citep{solheim2010,provencal1997}, exhibit dynamically stable mass transfer, although it has been suggested that some of these binaries, whose mass ratios are closer to unity, may undergo explosive events such as SNe~Ia or SN~.Ia \citep{bildsten2007a,perets2010a}.

In the AIC scenario, we assume that when any ONe\,WD approaches the Chandrasekhar mass limit via stable accretion, a NS is formed. Although {\sc StarTrack} can follow the evolution of each system after the formation of  the NS, this is beyond the scope of the present work to examine the future evolution of the entire AIC population.  However, we comment briefly on the future evolution of some AIC binaries in section\ref{sec:results} and section \ref{sec:discussion}.
We do not include AIC formed via CO WDs in our simulations, though in the Results we estimate the number of AIC NSs we expect could potentially arise from this channel. 

\section{Results}
\label{sec:results}
\subsection{Progenitors and delay times \label{ProgDelays}}

We find that AIC and MIC systems occur with a full range of delay times. Though both AIC and MIC NSs can be born up to a Hubble time (or beyond) after star formation, most are formed within 200\,Myr after starburst. 

\subsubsection{AIC donor types}

\begin{figure}
\includegraphics[width=0.5\textwidth]{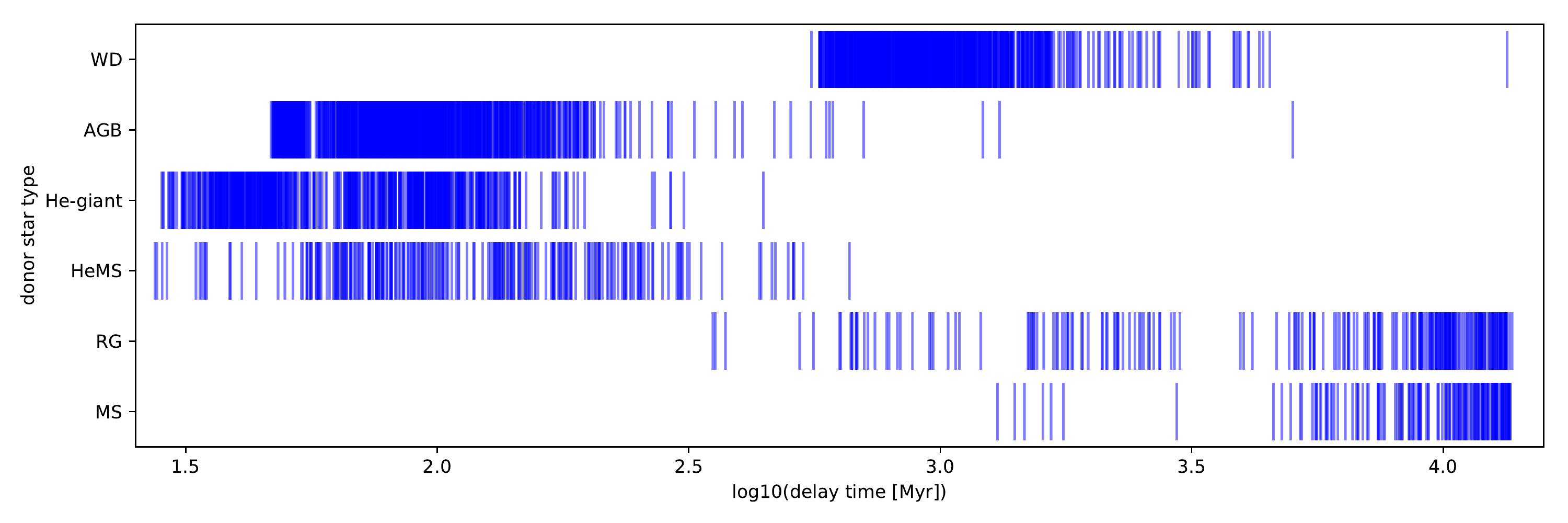}
\caption{Distribution of delay time vs. donor star type at time of AIC for the classical CE prescription. Each line represents one binary system. Regions of solid colour are regions of high number density in the delay time -- donor type parameter space.}
\label{dtddonstd}
\end{figure}

\begin{figure}
\includegraphics[width=0.5\textwidth]{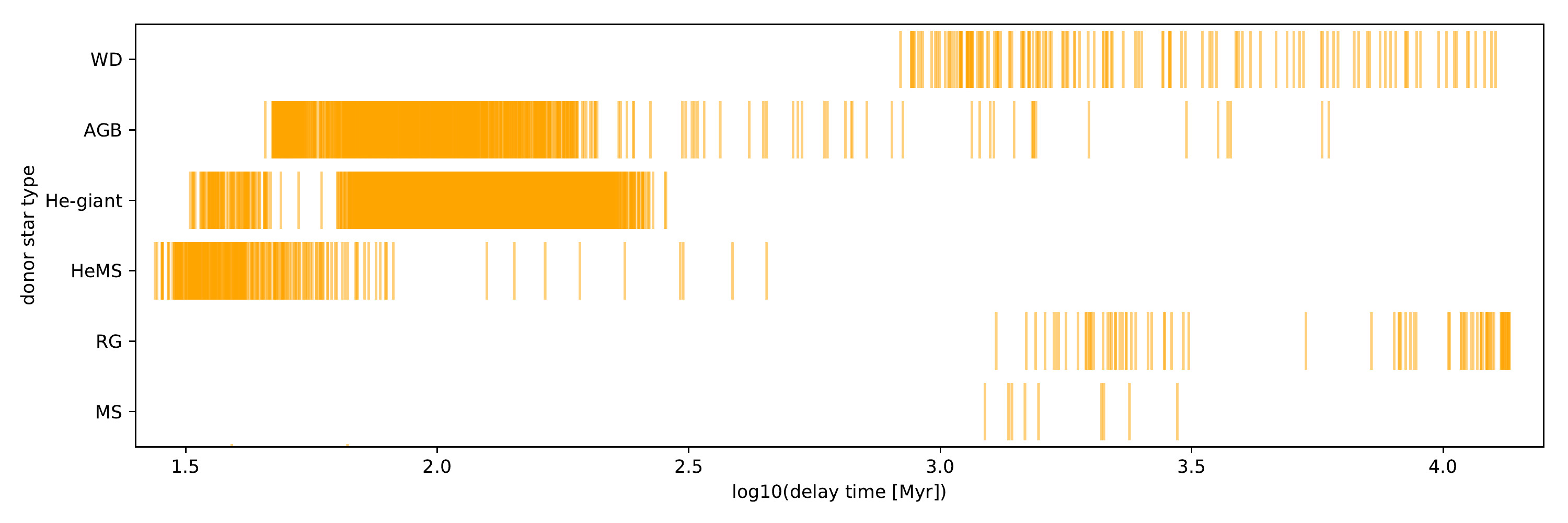}
\caption{Like Figure 1: distributions of delay time vs. donor star type at time of AIC for the new CE prescription.}
\label{dtddonnew}
\end{figure}

In Fig\,\ref{dtddonstd} and Fig\,\ref{dtddonnew} we show donor star companion type vs. delay time for the binary systems that undergo AIC for our classical and new CE models, respectively. The numbers are raw data and thus not calibrated to fit the characteristics (e.g. star formation rate over time) of the Milky Way. 
It is immediately obvious that AIC progenitors involving MS, giant-like or WD donors have the longest delay times; these are systems that plausibly represent the population of `young' NSs found among globular clusters. Their delay times start from greater than a few hundred Myr and extend to beyond a Hubble time. AIC progenitors with the most prompt delay times arise from binaries where the donor star has lost its hydrogen envelope but is still helium-burning (He giant or He MS). AIC progenitors that may arise via wind-accretion from a late AGB star have relatively short delay times, too (most within a few hundred Myr). 
The shortest delay time NS in our models is found in the New CE model, at a delay time of 28\,Myr with a stripped-envelope, compact helium-burning star donor (sometimes referred to as a main sequence helium star). The initial masses of the stars were 12.82 and 7.97\,\msun\, with an initial separation of 31\,\rsun.  We describe some specific evolutionary channels in the Appendix. 

Though the overall AIC birthrates are similar for both CE models (see section \ref{birthrates}), there are some notable differences between them, such as in the classical CE model, AIC progenitors with AGB donors dominate strongly over all other donor types (though comparable with H-envelope-stripped He-burning star and WD donors), while H-envelope-stripped He-burning stars strongly outnumber other donor types in the new CE model. The driving factor behind why different channels may be more prominent in one CE model but not in the other stems back to the fact that, in general, the new CE model gives rise to wider post-CE orbits. This is the reason why e.g. there is a factor of 6 more WD-AIC systems in the classical CE model vs. the new CE model. 

\subsubsection{MIC white dwarf masses}

\begin{figure}[t]
\includegraphics[width=0.5\textwidth]{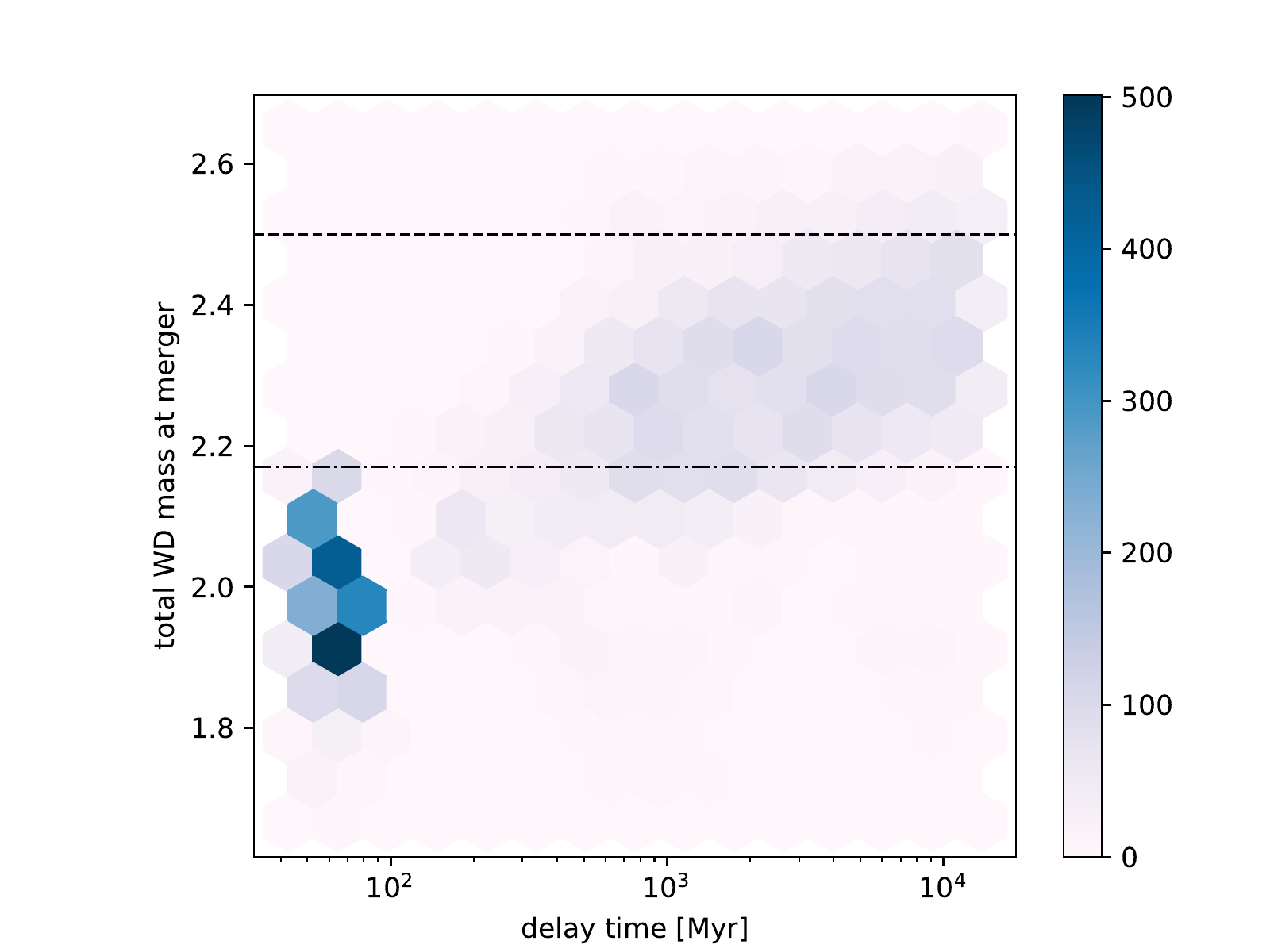}
\caption{Number density distributions of delay time vs. total WD mass at time of MIC for the classical CE prescription (darker colour means more systems in that time-mass bin). Horizontal lines correspond to the assumed neutron star upper mass limit, above which the compact object is likely to become a black hole in {\sc StarTrack} (dashed line) and \citet{margalit2017a} (dot-dashed line, see text).}
\label{dtdmtotstd}
\end{figure}

\begin{figure}
\includegraphics[width=0.5\textwidth]{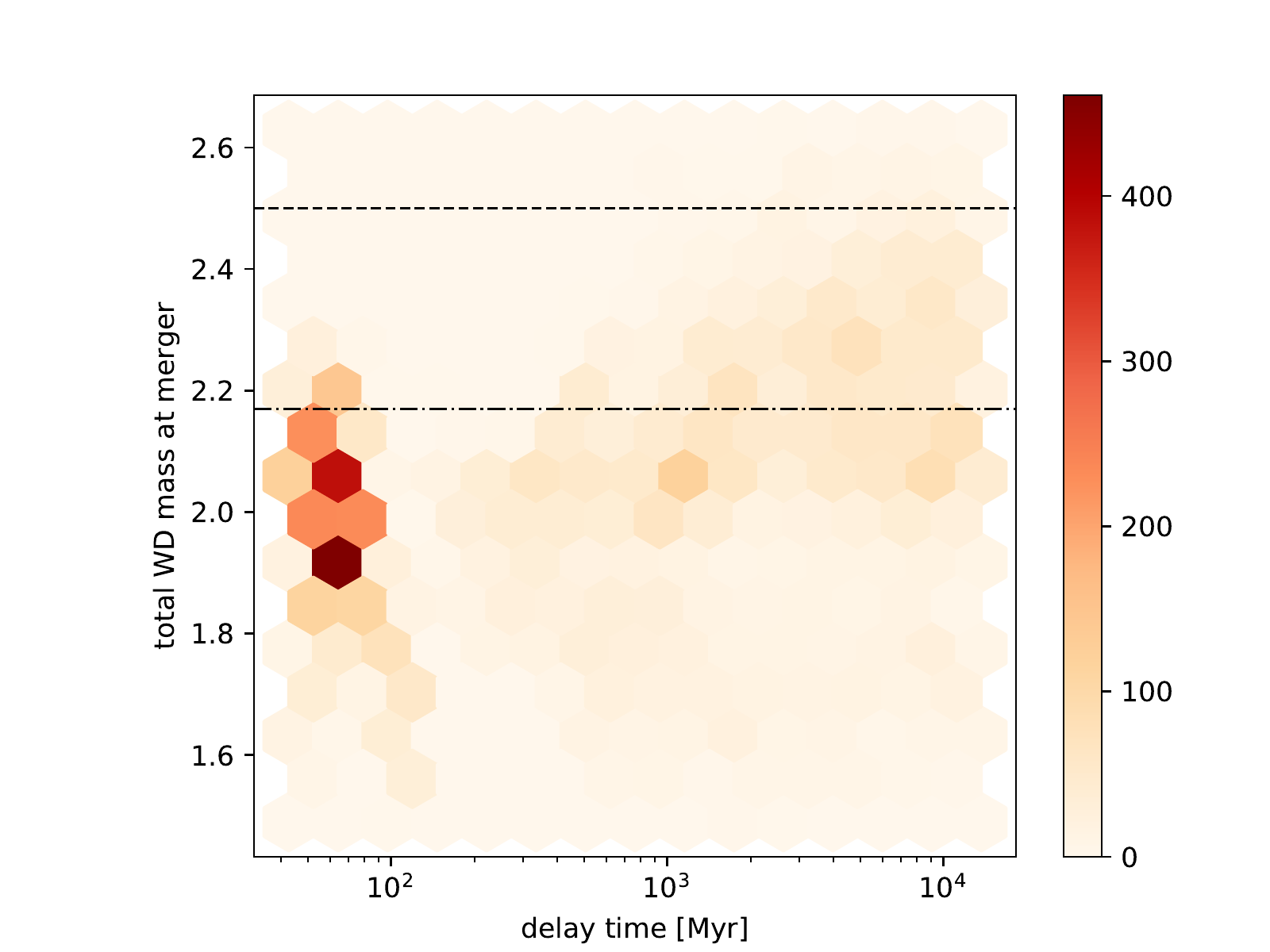}
\caption{Number density distributions of delay time vs. total WD mass at time of MIC for the new CE prescription. Horizontal lines correspond to the assumed NS upper mass limit, above which the compact object is likely to become a black hole in {\sc StarTrack} (dashed line) and \citet{margalit2017a} (dot-dashed line, see text).}
\label{dtdmtotnew}
\end{figure}

In Fig\,\ref{dtdmtotstd} and Fig\,\ref{dtdmtotnew} we show number density distributions of the total WD mass at time of MIC merger as a function of delay time for the classical and new CE models, respectively. 
Some massive binaries merge at late delay times for both CE models. These systems go through one CE event, rather than two CE events; the latter are common for the events with delay times $< 1$ Gyr. Note that in some systems it is the same star that undergoes a CE twice (first losing its hydrogen envelope, then later its helium envelope, see \citet{ruiter2013a}). Just like AIC NSs, MIC NSs can form at very long delay times.

We overplot with a dashed line the maximum NS mass that is normally allowed in {\sc StarTrack} ($2.5$ \msun\ is the typically-chosen parameter value). There are not many events above this threshold in either model, though we assume that in the event that a NS is born, any mass above this threshold would have to leave the system to avoid collapsing to a black hole. We also show with a dot-dash line the upper mass limit expected for a NS from the recent study of \citet{margalit2017a}, who find a maximum NS mass of only $2.17$ \msun. We find ${\sim} 50$  per cent and ${\sim} 20$  per cent of our MIC systems lie above $2.17$ \msun\ in our classical and new CE model, respectively. However, in the new CE (classical) model, about $90$ \%  (60 \%) of these compact mergers form through an evolutionary channel that involves an unstable phase of mass transfer while the mass-losing star is only a slightly-evolved giant. We assume that such a configuration would result in a CE phase, though other works have pointed out that binaries in this configuration may avoid a CE phase under these conditions (see Discussion).


\begin{figure}
\includegraphics[width=0.5\textwidth]{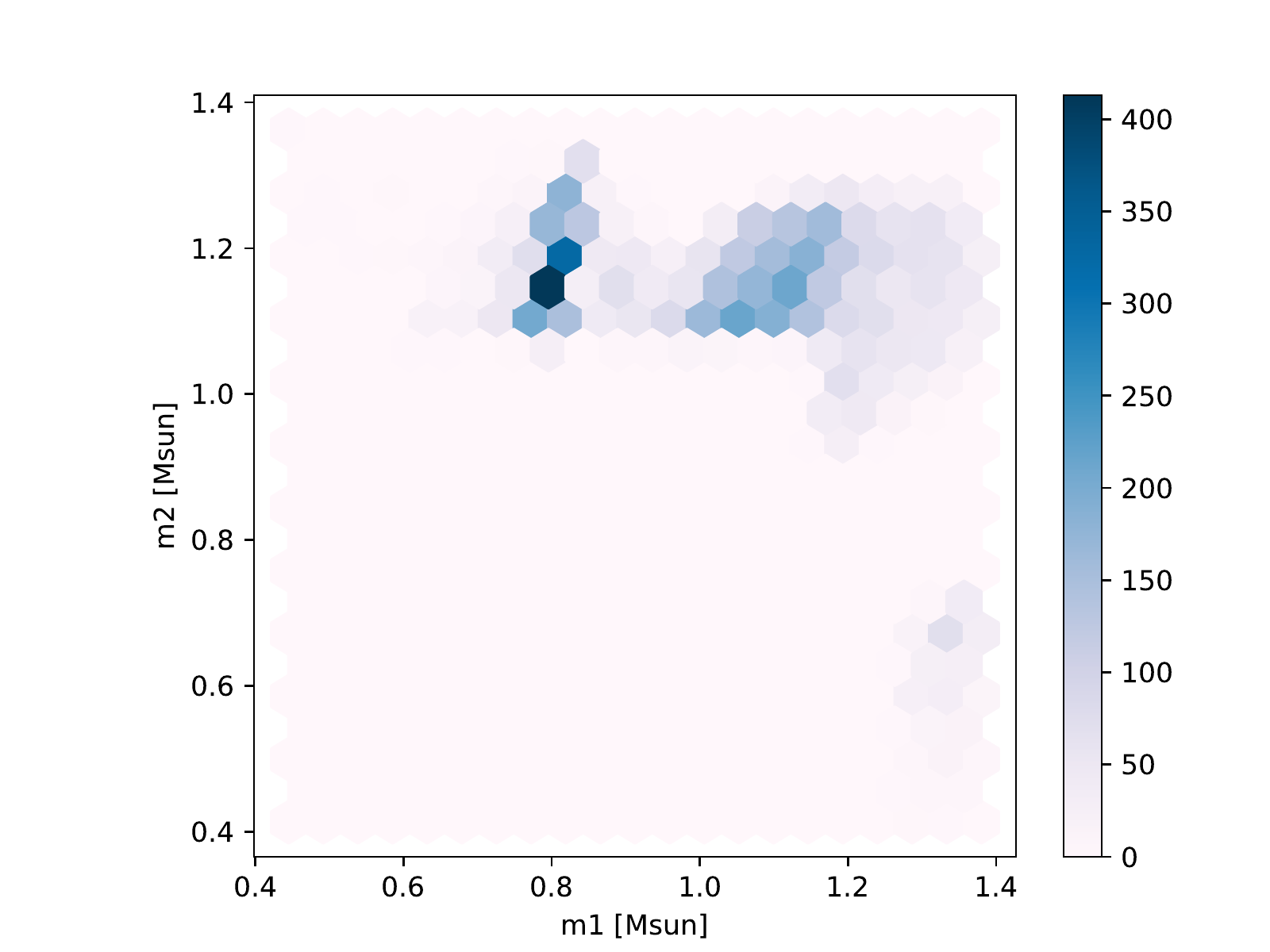}
\caption{Number density distributions of WD masses at time of merger for the initially more massive star (x-axis) and initially less massive star (y-axis) for the classical CE prescription (darker colours mean more systems in that mass-mass bin).}
\label{mstd}
\end{figure}

\begin{figure}
\includegraphics[width=0.5\textwidth]{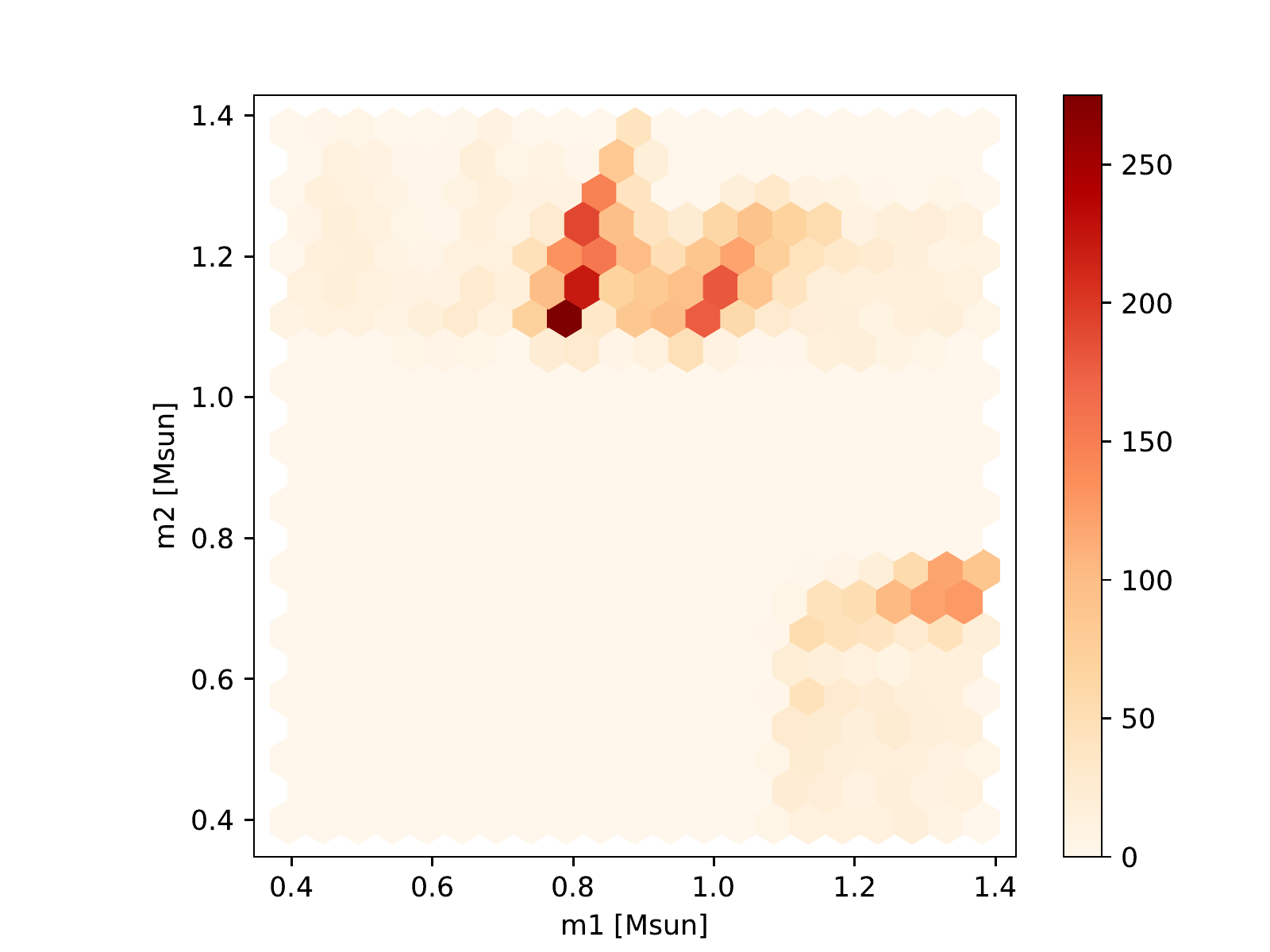}
\caption{Number density distributions for WD masses at time of merger for the initially more massive star (x-axis) and initially less massive star (y-axis) for the new CE prescription.}
\label{mnew}
\end{figure}

In Fig\,\ref{mstd} and Fig\,\ref{mnew} we show number density distributions of primary vs. secondary WD mass (initially more massive vs. initially less massive star). For comparison, a similar figure in \citet{lyutikov2017a} (see their figure 5) has a sharp discontinuity along the primary mass range ${\sim} 1-1.1$ \msun, though we note that they only consider CO+ONe WD pairs in their models (no double ONe WD pairs). It is not surprising that the different population synthesis codes reproduce similar results in terms of predicted pre-merger WD masses \citep[see also][]{toonen2014a}. The different regions of `clumps' is a result of different evolutionary channels (see sect. 3.3). 


\subsection{Birthrates\label{birthrates}}
We predict birthrates of AIC and MIC systems for our binary star populations. Our rates are calibrated per unit mass formed in stars and are calculated by assuming a binary star fraction of 70 per cent.
 There is no strong change in Galactic birthrate with adopted CE model. 
We find a current combined (AIC +  MIC) Galactic birthrate that is roughly an order of magnitude below the observationally-inferred Galactic SN~Ia rate: ${\sim}1.3 \times 10^{-4} (\pm 0.4 \times 10^{-4})$ and  ${\sim}1.7 \times 10^{-4} (\pm 0.4 \times 10^{-4})$ AIC per year for the classical and new CE models, respectively. We find a similar though slightly higher rate for MIC:  ${\sim}2.3 \times 10^{-4} (\pm 0.4 \times 10^{-4})$ and  ${\sim}2.3 \times 10^{-4} (\pm 0.4 \times 10^{-4})$, respectively. For our Galactic birthrates we assumed a constant star formation rate (see below).

The explanation for the lower number of ONe+ONe mergers in the new CE model is as follows. The main difference between the progenitors of ONe+ONe mergers and those of ONe+CO mergers is that ONe+ONe mergers typically undergo only one CE phase whereas CO+ONe mergers  normally undergo two CE phases. 
For the ONe+ONe MIC progenitors, the single CE event happens when the donor star is only slightly evolved (a red giant or in the Hertzsprung gap), so that it would still have a higher binding energy parameter $\lambda$ in the new CE model (by a factor of a few) compared to the classical CE model ($\lambda=1$; see section 2). As a consequence, the binding energy of the donor star in the new CE model is lower so the separation of the stars after CE is somewhat wider, and the binary never achieves contact again within a Hubble time. 
We comment on the likeliness of binary system survival during a CE phase when the giant-like donor does not have a strongly distinct core-envelope structure in the Discussion.


\begin{table*}
\begin{tabularx}{\textwidth}{|l | c | c | c | c |X |}    
\hline
Channel                          &  Class. CE tot & New CE tot &  Class. CE rate yr$^{-1}$ & New CE rate yr$^{-1}$  & general characteristic \\  
\hline
\hline
 AIC$+$main sequence          &     145    &   9     & $\leq$1e-5   &   $\leq$ 1e-6  & few \rsun $ \lesssim a \lesssim 10$ \rsun, donors $1-2$ \msun\ at AIC\\
\hline                                  
AIC$+$sub/red giant               &     231   &  81    & $\leq$ 1e-5     & $\leq$ 1e-6 & donors ${\sim} 1$ \msun\ at AIC, short post-AIC RLOF phase, end up as NS$+$He-rich WD binary on ${\sim}10^{1}-10^{2}$ \rsun\ orbit\\ 
\hline
AIC$+$stripped He-burning &     844   &  3178     &  3e-5  & 1.8e-4 & $1$ \rsun\ $ \lesssim a \lesssim 10$ \rsun, $\dot{M} {\sim}$few $10^{-6}$ yr$^{-1}$; post AIC LMXB phase with $\dot{M} {\sim}$few $10^{-8}$ yr$^{-1}$, later end up as NS$+$COWD binary on small orbit\\ 
\hline 
AIC$+$AGB {\em wind}            &   1112 & 1173     & 3e-5  & 3e-5  & thermally-pulsating AGB, wide orbit \\
\hline
AIC$+$white dwarf                   &  809    &   140    & 3e-5  & $\leq$ 1e-6  & $\dot{M} {\sim}$few $10^{-6}$ yr$^{-1}$,  usually He WD donor, then LMXB with $\dot{M} {\sim}$few $10^{-8}$ yr$^{-1}$\\
\hline
MIC CO$+$ONe                         & 7192  &    6017 & 2.2e-4 & 2.2e-4 & many channels; 1-3 RLOF phases and 1 or 2 CE phases\\
\hline
MIC ONe$+$ONe                     & 913     &   368     & 3e-5 & 4e-5 & usual channel involves 2 RLOF phases (H then He donor), 1 CE event (HG star), 1 RLOF event (He donor)\\
\hline
\end{tabularx}
\caption{Relative total number of AIC events (different donor channels) and MIC events from {\sc StarTrack} that occur per simulation of 12.8 million ZAMS binaries that are allowed to evolve for 13.7 Gyr. First two columns of values represent relative raw simulation AIC or MIC events over a Hubble time. The birthrate columns are expected current Galactic birthrates of AIC or MIC events. In the last column we list general characteristics typical of the progenitor (and in some cases post-NS evolution of the) system.}
\label{table:rates2}     
\end{table*}


In Table\,\ref{table:rates2}  we show relative formation rates (over a Hubble time) from our raw data for neutron stars formed through different AIC and MIC channels. We also present current birthrates estimated for the Milky Way Galaxy, assuming a constant star formation rate over 10 Gyr and a total stellar Galactic mass of $6.4\times 10^{10} $ \msun. The most dominant channel in our simulations is the hydrogen-stripped helium-burning star donor channel: we find similar results to \citet{wang2018a} for their CE model (our rates bracket, but are on the lower end of, the rates for two adopted $\alpha_{\rm CE}$ models of \citet{wang2018a}). However, our giant-like and main sequence donor channels are a factor of a few to an order of magnitude below those found by \citet{wang2018a}. 


For birthrates we have assumed the same simple star formation history of 6 \msun\,yr$^{-1}$ that was adopted in 
\citet{lyutikov2017a}. Our MIC birthrates are within the range found by \citet{lyutikov2017a}: ${\sim}2-5 \times 10^{-4}$ \msun \,yr$^{-1}$, depending on their adopted CE formalism.
Our predicted Galactic AIC rates are somewhat higher than those forecast by \citet{yungelson1998a} that amount to $8 \times 10^{-7} $ to $8 \times 10^{-5}$ events per year. 
When comparing our AIC birthrates to the binary millisecond pulsars formed via AIC of \citet{hurley2010a}\footnote{\citet{hurley2010a} also assumed a constant star formation history but at a higher rate of $8.6$\msun\,yr$^{-1}$, and used different assumptions for CE evolution.}, our numbers are within agreement (see their table 1) if we make the assumption that all AICs make MSPs. This is plausibly justified considering that conservation of angular momentum would naturally cause the NS to spin at millisecond periods and the conservation of magnetic flux would produce a NS with magnetic fields in the range $10^8-10^9$\,G, typical of MSPs, without having to invoke accretion induced field decay \citep{ferrario2007a,smedley2015a}. On the other hand, \citet{tauris2013a} pointed out that the formation of NSs  may be inevitably accompanied by the generation of strong magnetic fields, and thus these AIC NSs would need to undergo further accretion to reach the observed weaker fields. 

As mentioned previously, it is theoretically possible that a CO WD can undergo quiescent burning and evolve into an ONe WD under the right burning conditions. It is therefore worthwhile to estimate how many accreting CO WDs in our {\sc StarTrack} models may contribute toward the AIC progenitor population. In {\sc StarTrack}, the canonical assumption is that if a CO WD approaches the Chandrasekhar mass limit it will produce a Type Ia supernova, regardless of the accretion rate, but we have estimated the number of AIC NSs that could arise from the CO WD channel in a post-processing step.  Following \citet{wang2017a}, we check our simulations for CO WDs that undergo accretion where the initial accretion rate exceeds $2.05 \times 10^{-6}$ \msun yr$^{-1}$ and the CO WD mass is close to Chandrasekhar ( $> 1.35$ \msun) at the end of the accretion phase. We find that the CO WD AIC channel can contribute up to 20\% of what we find from our combined ONe AIC WD channels. This fraction is in rough agreement with the results of \citet{wang2018a}. 

We note that our predicted Galactic rates of AIC events are in agreement with upper limit estimates set by solar system abundances of neutron-rich isotopes, which is ${\sim} 10^{-4}$ AIC per year  \citep{fryer1999a,dessart2006a,metzger2009a}. Though it remains to be confirmed which isotopes are synthesized in AIC events, if AIC NSs are indeed a formation site for a sub-set of neutron-rich isotopes (including r-process), local abundance patterns could potentially provide a useful lower limit on Galactic AIC formation rates \citep{desilva2015a}. 
If we consider the fact that AIC NSs may leave behind a hot remnant of ionised gas that is potentially visible for up to 100,000 years, analogous to what has been predicted for the Chandrasekhar mass (i.e. single degenerate) scenario of Type Ia supernovae \citep{woods2017a}, we estimate that on the order of ${\sim}$10 of these post-AIC remnants could currently exist in the Galaxy today.

\begin{figure}
\includegraphics[width=0.5\textwidth]{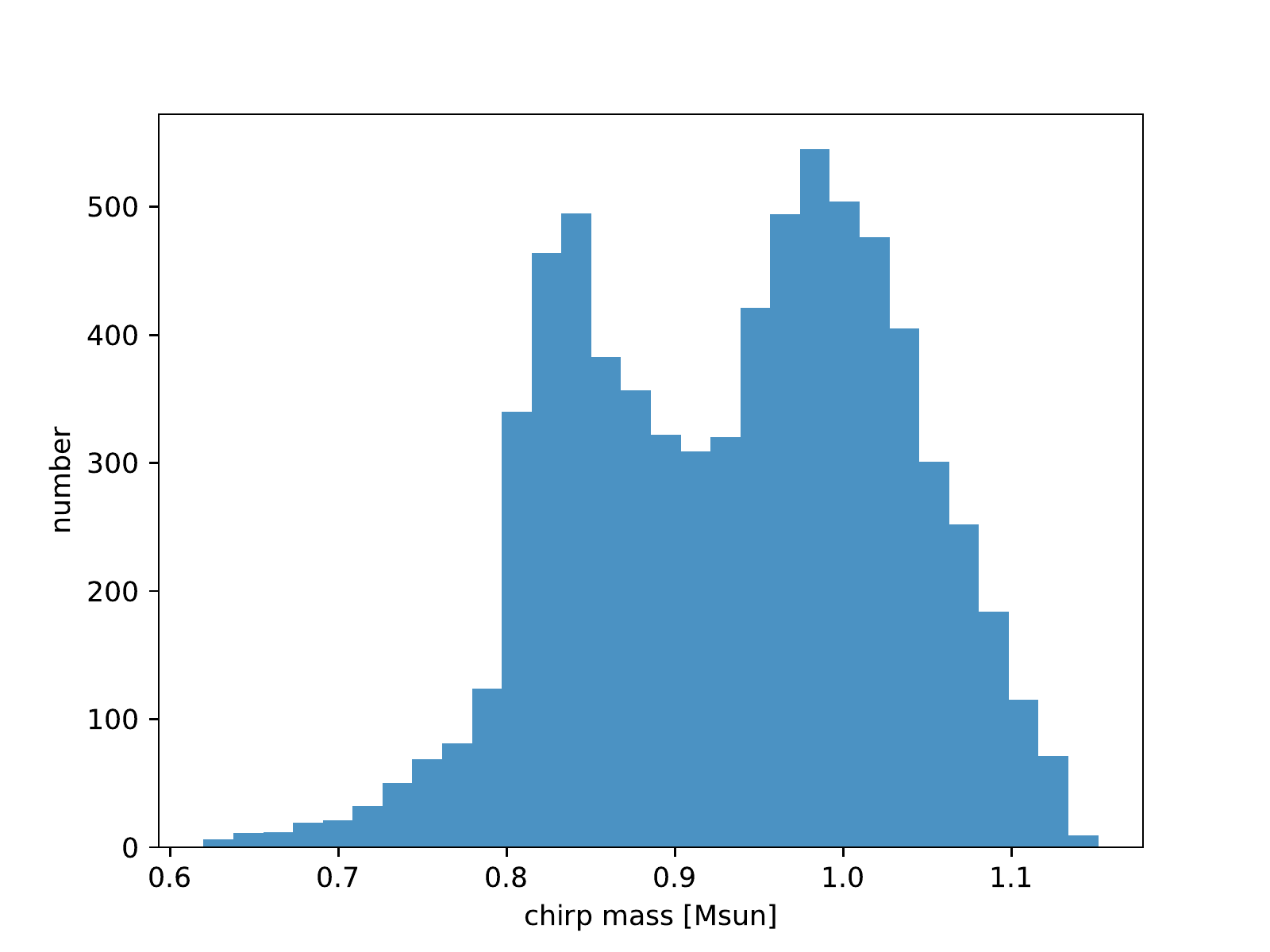}
\caption{Chirp mass  of white dwarf pairs leading to MIC events for the classical CE model.}
\label{chirpstd}
\end{figure}
\begin{figure}
\includegraphics[width=0.5\textwidth]{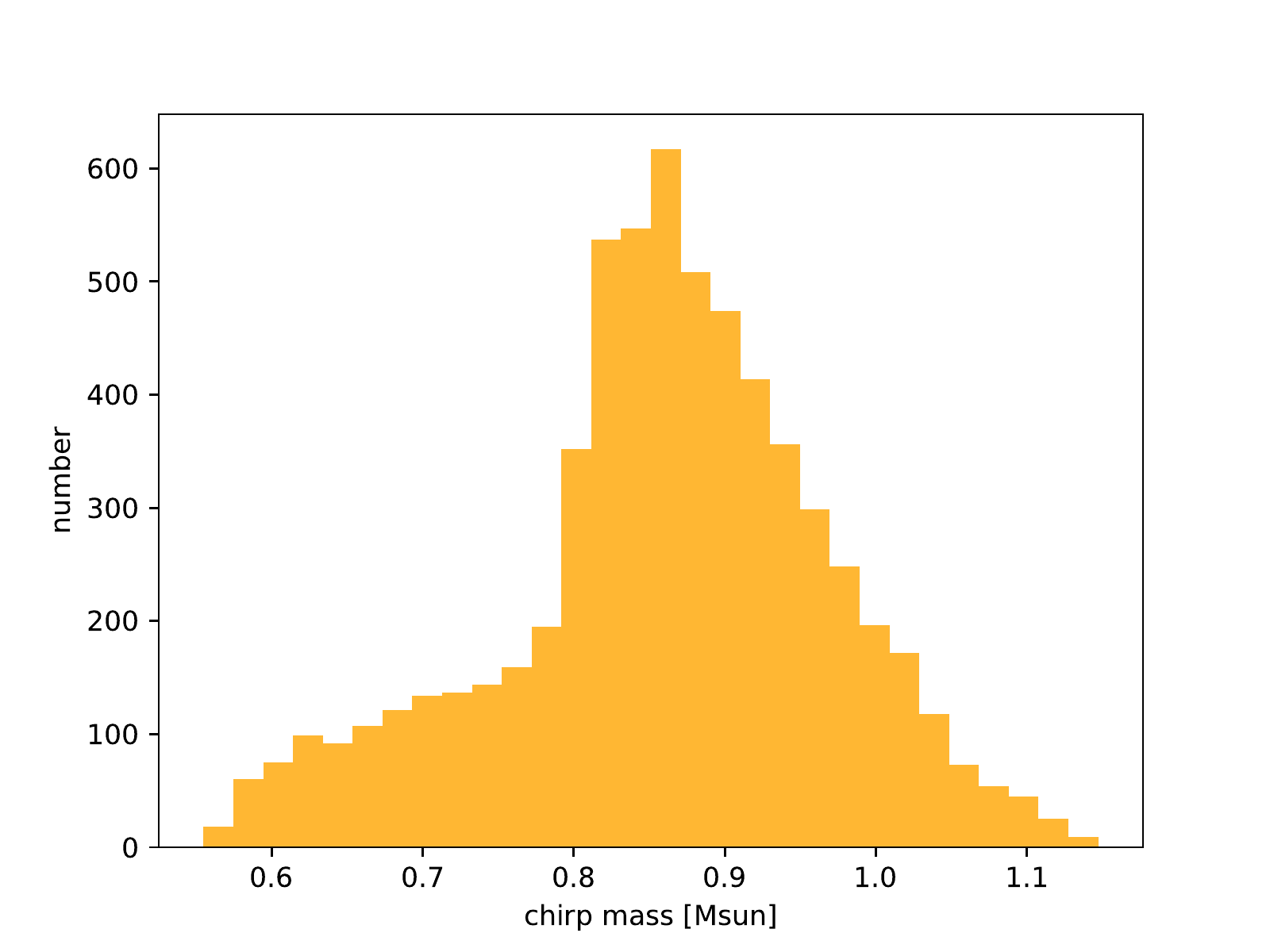}
\caption{Chirp mass distribution of heavy white dwarf pairs for the new CE model.}
\label{chirpnew}
\end{figure}

\begin{figure}
\includegraphics[width=0.5\textwidth]{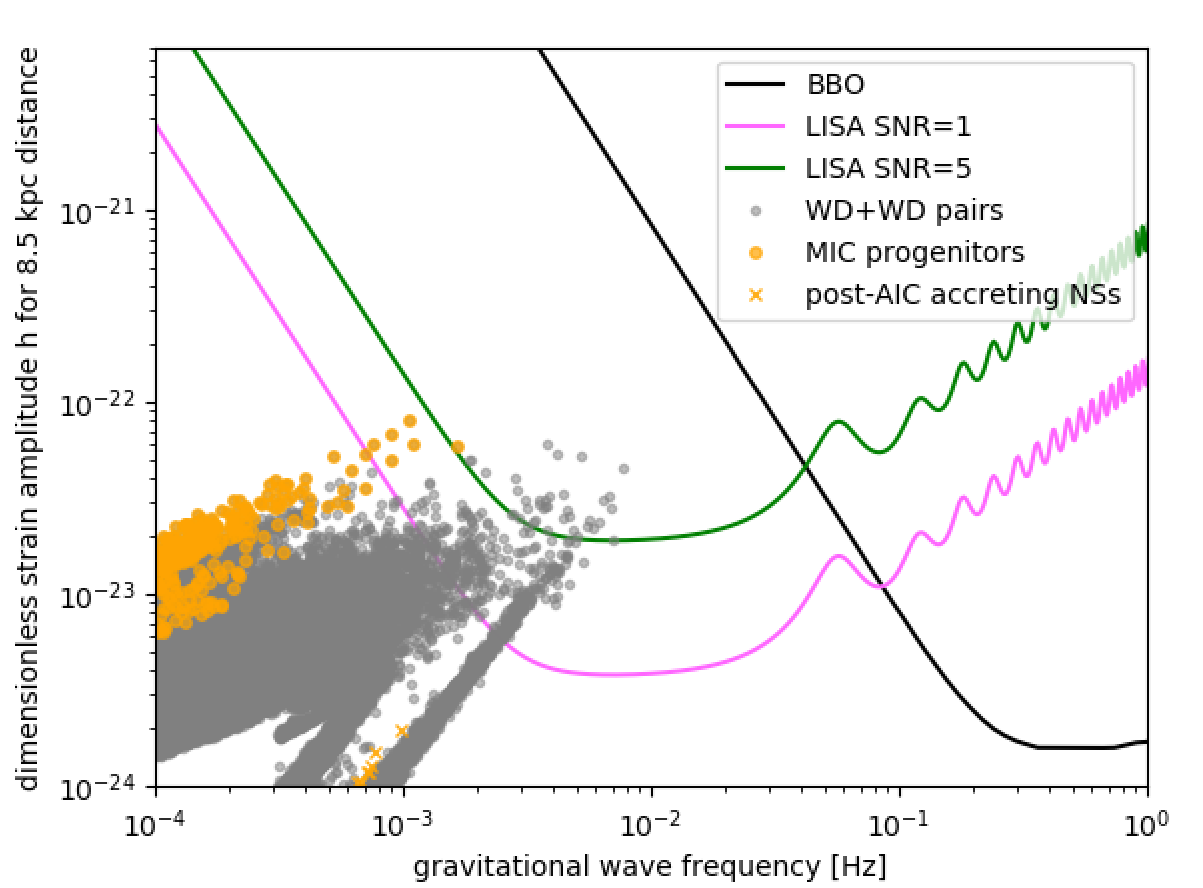}
\caption{Gravitational wave strain amplitude of Galactic double white dwarfs (grey), including MIC progenitors (orange dots) as well as post-AIC binaries that have since started a phase of RLOF (e.g. low-mass X-ray binaries; orange crosses) for our new CE model. We have plotted the strain amplitude for a distance of 8.5 kpc to give a general idea of the parameter space (in terms of h and f$_{\rm gw}$) that we expect our binaries to populate at the current epoch. Note that the true gravitational wave signal will depend on several factors, distance being one of the most important. We over-plot the sensitivity curve for the proposed future space-based gravitational wave observatory LISA (signal to noise of 1 and 5 in magenta and green, respectively, from Shane Larson's sensitivity curve plotter), and BBO from \citep{yagi2017a}.}
\label{fgw-current}
\end{figure}

\subsection{MIC and AIC progenitors as gravitational wave sources}
\label{sec:GW}

It has been shown that the gravitational wave signature of Galactic double WD binaries with orbital periods of less than about 5 hours will be detectable with future space-based gravitational wave observatories, like the Laser Interferometer Space Antenna, LISA \citep{nelemans2004a,ruiter2010a,cornish2017a}. The most massive, very nearby binaries with the shortest orbital periods  will have the best chance to be individually resolved as gravitational wave sources \citep{nelemans2013a}. 
A quantity that is a function of the component masses, the chirp mass $M_{\rm chirp} = (m_{1}m_{2})^{3/5} / (m_{1} + m_{2})^{1/5}$, can be directly measured from gravitational wave signals. The (dimensionless) strain amplitude, $h$, is proportional to $M_{\rm chirp}^{5/3}$, via

\begin{equation}
h = 2 (4\pi)^{1/3} \frac{G^{5/3}}{c^{4}} f_{\rm GW}^{2/3} M_{\rm chirp}^{5/3} \frac{1}{r}
\end{equation}
where the gravitational wave frequency of a binary with a circular orbit is  $f_{\rm GW}=2/(P_{\rm orb})$ where $P_{\rm orb}$ is the orbital period, and $r$ is the distance to the binary from the detector. $h$ is a measure of the fractional change in separation that occurs (e.g. between detector nodes, or spacecrafts) when a gravitational wave passes through the detector. Naturally, as the binary approaches merger time, its orbital period decreases drastically and thus $f_{\rm GW}$ increases (sweeping, or `chirping' across the frequency band). Note that the orbital frequency evolution can also be negative, such as in the case of some mass-transferring binaries \citep{kremer2017a}.

In Fig~\ref{chirpstd} and Fig~\ref{chirpnew} we show the distribution of MIC progenitor chirp masses for the classical and new CE approaches, respectively. 
The distribution of chirp masses is bimodal in the classical CE model, but it has only one peak in the new CE model. The reason for the different distribution shapes is attributed to the different formation channels that dominate depending on the CE model. The fact that different formation channels are more (less) prominent than others depending on the CE model results in a different ranges of chirp mass being more (less) populated than others.

In the classical CE model, two main peaks are visible in Fig~\ref{chirpstd}: the highest peak is ${\sim} 0.83$ \msun, and another broader peak centered ${\sim} 1$ \msun\ (see Example 2 in the Appendix). The first peak corresponds to progenitors with delay times less than 300 Myr arising from different formation channels. The systems with chirp masses below ${\sim} 0.79$ \msun\ are almost exclusively arising from progenitors with delay times greater than 300 Myr that undergo two CE events: first when the primary is an AGB star, then later when the secondary is a red giant. These MIC systems populate the lower-right quadrant of Fig. 5. 

In the new CE model for MIC progenitors, the peak in Fig~\ref{chirpnew} ${\sim} 0.87$ \msun is dominated by two formation channels: one short (${\sim}70$ Myr) delay time channel (see Appendix example 3) and one long (${\sim}11$ Gyr) channel. The systems that populate the parameter space with lower chirp masses consist mostly of short delay time MIC binaries that undergo one stable mass transfer event (primary star loses mass to secondary), and two CE events (primary and then secondary star losing envelope, or secondary star loses its envelope: first hydrogen envelope then later the helium envelope). These MIC systems populate the lower-right quadrant of Fig. 6. The binaries populating the higher chirp mass part of the distribution are almost exclusively from one formation channel with very short (${\sim} 40$ Myr) delay times and the following mass transfer phases: 1. stable mass transfer from a Hertzsprung Gap star (primary) toward main sequence star, 2. stable mass transfer from a naked helium-giant (primary) to a main sequence star, 3. a CE phase between a (secondary) star in the Hertzsprung Gap and a CO WD, and 4. stable mass transfer between a H-envelope-stripped helium-burning giant (secondary) and the CO WD. This final phase is the higher mass analogue of what occurs during the evolution of some Type Ia supernova progenitors (stage VIII) uncovered in \citet{ruiter2013a}. 

If binaries in fact do survive a CE phase that involves a Hertzsprung Gap star (see section \ref{sec:discussion}), then our predictions indicate they will have some of the highest chirp masses among the double WD binary population. Given the relatively low birthrate of NSs created by MIC, we do not expect to `see' a MIC event with LISA (would have to be lucky!), but our calculations indicate that a number of MIC progenitors will be visible in the LISA band (see below). Since MIC progenitors comprise a sub-population of the total double WD population, we can be more optimistic at the prospect of detecting the lower-mass (and more numerous) double WD systems, e.g. CO$+$CO WD binaries, many of which are likely Type Ia supernova progenitors. We look forward to the time when gravitational wave measurements of close binary stars can finally confirm or refute different ideas about what drives CE evolution, thus providing critical insight into CE physics. 

It is certainly worth noting that, regardless of how the CE phase behaves in Nature, MIC progenitors should be detectable in gravitational waves, in fact they could even be resolved,  both in the Galaxy and in the Magellanic clouds. Though their birthrates will likely be lower in the Magellanic Clouds than in the MW owing to the smaller stellar mass of the LMC and SMC, the fact that the LMC and SMC are more actively star-forming (per unit stellar mass) means that the MIC birthrates may not be too far off from those we predict for the MW.  

In Fig~\ref{fgw-current} we show the gravitational wave strain amplitudes for different binary systems in our new CE model, noting that these results do not change significantly with adopted CE model. In this work we have not attempted to replicate a sophisticated model of the Milky Way in terms of star formation history and stellar density distribution. In reality, the gravitational wave signal will depend on not only the physical properties of a binary (as well as its distance), but will also involve assumptions about the instrument (and noise) itself. 
We plot the expected sensitivity curve for the future space-based gravitational wave observatory LISA for two signal to noise ratios and the expected sensitivity curve for the Big Bang Observatory (BBO). 
As a first order approximation to the expected strain amplitude imparted by the population of MIC and AIC sources, we plot the strain amplitude for various binary systems with orbital periods less than 5.5 hrs at the current epoch assuming a distance of 8.5 kpc. We have again assumed a constant star formation history for 10 Gyr and a MW age of 10 Gyr with a total Galactic mass born in stars of $6.4 \times 10^{10}$ 
\msun.\footnote{We note that the results shift up or down by a factor of ${\sim 10}$ and ${\sim 8}$, respectively in terms of $h$, when distances of 1 kpc and 50 kpc (LMC distance) are assumed. 

We show the entire simulated double white dwarf population with grey circles and the MIC progenitor sub-population is overlaid with orange circles. In orange crosses we show {\em post-AIC} binaries that are undergoing a LMXB phase, having since initiated a phase of RLOF between a WD and NS. These plausible electromagnetic counter-parts to post-AIC LISA binaries populate a similar region of parameter space as other mass-transferring double WDs, e.g. the AM CVn systems. In our simulation data, we do not find any post-AIC LMXBs that would currently be visible with LISA at the assumed distance (8.5 kpc), though the binaries with the highest gravitational wave frequencies (shortest orbital periods) start to be potentially visible around the SNR$=1$ curve when a distance of 1 kpc is assumed. We also note that Figure 9 represents data from a simulation where the total simulated stellar mass was $1.63 \times 10^{8}$ \msun, which is a factor of $392$ lower than the Galactic stellar mass. 
Due to the fact that we have not simulated the total number of stars in the Galaxy, combined with Monte Carlo effects, it may indeed be possible that a few to tens of post-AIC LMXBs could be detectable with LISA. An example of such a {\sc StarTrack} binary that appears with the highest gravitational wave frequency (one of the orange `x' symbols in Fig. 9) is a HeWD+NS system with an orbital period of 34 minutes and estimated X-ray luminosity of ${\sim} 1 \times 10^{38} $ erg s$^{-1}$.  
}


\section{Discussion}
\label{sec:discussion}

As mentioned previously, some formation channels require the CE phases to occur while the mass-losing star is only a slightly evolved giant, e.g. in the Hertzsprung Gap. However, the assumption that a binary can survive such a CE phase is rather optimistic. Stars in the Hertzsprung Gap have a more poorly-defined core-envelope structure as compared to more evolved red giants or AGB stars that have a well-defined jump in specific entropy between the core and envelope \citep{belczynski2007a}. Further, stars that we have found to undergo significant mass loss (and initiate a CE phase) while in the Hertzsprung Gap may rather experience stable mass transfer instead \citep{pavlovskii2015a}, which would change the evolutionary outcome of the binary altogether, likely lowering the number of potential MIC systems arising from certain formation channels. 

As noted in the Results, we do not produce any AIC NSs via the formation channel where a CO WD is produced first, then evolves into an ONe WD via accretion and subsequent carbon shell flashes, though we estimate the  contribution from this channel to be on the order of 20\%.  
This COWD-to-AIC scenario was discussed in \citet{brooks2017a} \citep[see also][]{wang2018a}, and \citet{brooks2017a} state that this may be a significant formation channel leading to AIC events. In our simulations, all AIC progenitors of ONe WDs have masses above 7 \msun\ on the ZAMS and, at the base of the AGB, their cores are above the minimum value (threshold between forming CO core and ONe core) that is needed to form an ONe WD (in {\sc StarTrack}). Our models predict some AIC systems that are similar to those considered in \citet{brooks2017a} (a H-envelope-stripped helium-burning star donor with a mass ${\sim 1.5}$ \msun\ when the final phase of pre-AIC mass transfer). The following example is found in our new CE model: The AIC progenitor's evolutionary channel consists of several mass exchange events (including two CE events) with the final pre-AIC RLOF phase (between the H-envelope-stripped helium-burning star and the ONe WD) starting 47 years after star formation. The AIC occurs just 0.5 Myr later (delay time 47.5 Myr). 
The NS+He star binary survives to later briefly re-establish stable RLOF between the H-envelope-stripped helium-burning star and the NS, (becoming a low-mass X-ray binary candidate), which continues even as the donor evolves off the helium main sequence \citep[see also][]{liu2018a}.  RLOF ceases at 61 Myr with component masses 1.44 and 0.71 \msun, respectively, for the NS and evolved naked helium star. The helium star soon thereafter evolves into a CO WD and at this time the system has an orbital period of 125 minutes. Emission of gravitational radiation causes the NS-WD binary to merge 149 Myr after star formation.

We find that in our simulations, AIC neutron stars that form at long delay times can exhibit properties  that match those of so-called "redback" MSPs fairly well. 
Redbacks are binary MSPs with orbital periods shorter than about 1.5\,d, low to zero eccentricities, and the pulsars' companions have minimum masses estimated to be in the range ${\sim} 0.1-0.5$\,\msun\ \citep{smedley2015a,manchester2005a}. 
These systems exhibit long radio eclipses caused by circumbinary discs and do not lie on the $M_c−P_{orb}$ relation for MSPs that would be expected from the evolution of red giant donors later cooling to become He\,WD 
companions. \citet{smedley2015a}  proposed that redbacks and some black widow pulsars could be the result of the AIC of an ONe\,WD. In {\sc StarTrack}, we find that those AIC systems that end up as non-accreting AIC NS+WD binaries with low-mass WD companions and short orbital periods (resembling observed redback systems) arise from different progenitor channels depending on the adopted CE treatment. 
Binaries exhibiting redback-like properties undergo AIC when the donor is on the Main Sequence, red giant branch or a helium WD in the classical CE model, and when the donor is a helium white dwarf  in the new CE model. 
Our current study clearly supports the (long delay time) AIC route for the formation of this exotic class of MSPs. 

For over 80 per cent  of AIC systems from our models, the companion star has evolved into a WD by a Hubble time. In cases where the companion is not a WD by a Hubble time it is almost always a H-envelope-stripped, helium-burning star (e.g. a CO WD progenitor). Observations \citep[see][]{manchester2005a} show that nearly 50  per cent of binary MSPs (about 60  per cent of the total population) have a helium-WD companion, 10  per cent a CO\,WD, 10  per cent are redbacks and about 20  per cent have an ultralight companion (a substellar object or planet). Thus our synthetic AIC population is in general agreement with current observations of MSPs, while the NSs forming via MIC will contribute to the population of isolated MSPs.
We also note the formation of a few widely separated (though still bound) double NS binaries (separations of about $10^{4}$ \rsun). In this case both ZAMS masses were in the narrow range $7.5 - 7.8$ \msun. 

\subsection{Impact of initial orbital configuration}

We briefly discuss the general impact that adopting different assumptions for initial orbital distributions has on our results. Adopting the new CE model, we performed additional simulations adopting two different sets of initial distributions for orbital parameters. For one model we used a thermal eccentricity distribution as adopted in \citet{ruiter2009a}, and in another we adopted the initial orbital period, eccentricity and mass ratio distributions described in \citet{sana2012a}. Hubble-time averaged AIC rates increased by $16$\% when adopting the thermal eccentricity distribution, and decreased by $19$\% when the \citep{sana2012a} distribution is adopted. For the MIC systems, in both cases the Hubble-time averaged rates increased by 37\% and 78\% respectively. 

A contributing factor to the boost in numbers of very close binaries (that will merge) in the model with a thermal eccentricity distribution is the fact that, for a small population of binaries that are initially eccentric, the orbit will not have circularised from tidal interactions before the first mass transfer phase takes place. Evolutionary phases in {\sc StarTrack} that involve ongoing mass transfer assume circular orbital configuration, so for the small number of binaries that have not circularised by the time RLOF starts we set the binary separation to periastron position, thereby decreasing the orbital semi-major axis (a similar system but with $e0=0$ would not ever get close enough to merge in a Hubble time). To an extent this effect also plays a role for the \citet{sana2012a} model, but the assumptions for that model are quite different from our adopted model and the physical dependencies are more complex and harder to disentangle. 
The task of unravelling all of the outcome dependencies of how various assumptions alter the progenitor population is beyond the scope of this particular work, and such a parameter study would be a paper on its own \citep[e.g.][]{claeys2014a}. It is worthwhile noting however that the total number of white dwarf mergers is lower for the adopted model compared to the other two considered test models (this also applies to SN Ia progenitor candidates, e.g. CO+CO WD mergers, not just the heaviest WD mergers). Thus our computed MIC birthrates could be seen as a lower limit - especially considering the fact that we have not included any double COWD binaries in our MIC calculations.

\subsection{Globular clusters and the influence of initial helium fraction}

We find that induced-collapse NSs can be born at any delay time ranging from $\sim 30$ Myr to a Hubble time and beyond (see section \ref{ProgDelays}).   
It has long been postulated that a low kick velocity channel for the production of NSs may be required to explain the large excess of these objects in globular clusters \citep{katz1975a}. 
Core-collapse supernovae may result in large natal kick velocities being imparted upon their newly-formed NSs, owing to asymmetries that arise in mass ejection and/or neutrino emission \citep{janka2017a,fryer2006a}. 
MSPs derived from AIC are expected to suffer much smaller natal kicks than delivered to NSs resulting from core-collapse supernovae \citep[e.g.][and references therein]{freire2014a}. This may help to explain how, despite their shallow gravitational wells, so many MSPs can be present in globular clusters \citep{grindlay1987a,bailyn1990a,benacquista2001a,podsiadlowski2004a,ivanova2008a,freire2013a}. 

It is a worthwhile exercise to consider how different chemical environments can influence the formation of compact objects. In particular, enhancement of helium on the ZAMS can have a significant effect on the final evolution of the star -- whether it evolves into a CO WD, ONe WD, or NS \citep{shingles2015a}. 
HST photometry has revealed that some metal-poor Galactic globular clusters host extremely helium-enhanced stellar populations, with helium abundances up to $Y \approx 0.4$ \citep{gratton12}. 
An increase in helium for the Galactic bulge has also been suggested based on observations of the red giant branch bump \citep{nataf11,bensby13} and the discrepancy between its photometric and spectroscopic turnoff ages \citep{nataf12}. Helium-enhanced sub-populations have also been suggested for early-type and spiral galaxies \citep{atlee09,chung11,rosenfeld12,buell13}.

While the effect of helium-enrichment on stellar evolution is relatively well known e.g., the main sequence luminosity will be higher owing to an increase in the mean molecular weight, most studies have focused on low-mass stars and the effect of helium enrichment on colour-magnitude diagrams \citep[e.g.,][]{sweigart87,salaris05,chantereau15}. Single star models of intermediate-mass that are evolved past the main sequence also show interesting behaviour, with helium-enhanced models entering the AGB with a more massive H-exhausted core compared to their primordial helium counterparts \citep[e.g.,][]{karakas14a}. The main consequence from that study is that the minimum mass for carbon burning is reduced from $> 6 $\msun\ for $Y=0.24$ to $4-5$\msun\ for $Y = 0.35 -0.40$ for [Fe/H] $\approx -1.4$  \citep{shingles2015a}. 

In our simulations we have adopted a canonical value for helium of $Y=0.28$, similar to \citet{karakas14b}. Interestingly, very metal-rich models of $Z=0.1$ also show a decrease in the minimum mass for carbon burning to $\approx 7$\msun\ compared to solar metallicity, a result which will applicable to very metal-rich stellar populations including those found in massive early-type galaxies. 


Because the stellar initial mass function \citep{kroupa1993a} favours the production of lower mass stars, the fact that the carbon-burning mass threshold can be decreased by up to two solar masses would have noticeable consequences. 
Larger initial helium abundances yield a larger number of ONe WDs and correspondingly, fewer CO WDs. Thus the rate of SNe\,Ia from merging CO WD pairs would be decreased, as would the rate of SNe\,Ia involving RLOF toward a CO WD. At the same time, we would expect the rate of ONe production, and thus the AIC and MIC rates, to increase from the predicted Galactic rate of a few $\times 10^{-4}$ per year. This proposed effect might be somewhat curtailed if CO WD mergers on the lower mass end, e.g. even those WD pairs with combined masses below the Chandrasekhar mass limit, are an important contributor to SNe~Ia \citep{vankerkwijk2010a}. However, even though including the lower-mass WD systems as potential SN\,Ia progenitors brings the theoretical rate predictions into agreement with observations \citep{badenes2012a,maoz2018a}, the idea is still somewhat speculative since it remains to be demonstrated how the explosion would be triggered, and what the burning products might be. 


\section{Summary}
\label{sec:summary}

We have predicted birthrates, delay times, and progenitor configurations for NS production in binary star systems that contain at least one massive WD. We find that our AIC and MIC rates, calculated with the {\sc StarTrack} code, are in agreement with those found by \citet{hurley2010a} (BSE code) and \citet{lyutikov2017a} (SeBa code), respectively. 
Our present work substantiates the need to consider NSs formed via induced collapse as an important channel for NS formation.
In the future, when WD binaries can be resolved through gravitational radiation, measuring their chirp masses will be possible, and with enough statistics these binaries will be useful in constraining their own evolutionary origin. As we have shown, it may be possible to set limits on CE physics by determining chirp masses of double WDs, since the two CE prescriptions adopted here exhibit distinct $M_{\rm chirp}$ distribution shapes (Figs. 7 and 8). While MIC events are not expected to be detectable with LIGO, telescope array surveys like BlackGEM should be able to detect optical counterparts of merging double white dwarfs beyond our Galaxy. Such observations would be very useful for the successful gravitational wave detection of MIC binaries with LISA.

Particularly interesting results of our study are  1) a number of nearby merger-induced collapse progenitor systems should be observable with sensitive space-based gravitational wave detectors such as LISA 2) a smaller number of post-AIC systems that are undergoing a phase of RLOF between a WD and a NS could be observable with LISA and moreover should have electromagnetic counterparts as LMXB sources 3) induced-collapse NSs can be born at extremely early times after the onset of star formation, e.g. 30\,Myr, as well as at extremely late times (Hubble time), and 4) in the AIC scenario, different types of donor stars are responsible for pushing the accreting WD toward the Chandrasekhar mass limit, and except for the AGB-wind donor case, AIC delay times reflect the evolutionary timescale of the donor. In particular, if we allow a WD to accrete via a wind from an evolved (AGB) star, the AIC birthrates are noticeably enhanced, and a substantial fraction of these NSs are born with delay times $< 100$ Myr (Figs 1 and 2). These AGB donor systems should be relatively bright, wide binary systems, but given the short-lived (${\sim} 1$ Myr) evolutionary phase and the current Galactic rate estimate for the AGB channel ($3 \times 10^{-5} $ per year), we only expect the MW Galaxy to currently harbour ${\sim}$30 such systems. The hydrogen-stripped, helium-burning star donor channel, on the other hand, has an estimated Galactic birthrate of nearly $2 \times 10^{-4}$ per year in the new CE model.  The RLOF phase prior to the AIC event is short-lived ($\lesssim 0.1$ Myr), but the post-AIC, LMXB phase lasts on the order of 1 Myr.  Thus, we estimate on the order of 200 such post-AIC accreting NS systems to exist in the Milky Way, some of them potentially resolvable with LISA. If the duty cycle of such LMXBs is as high as 10\%, then ${\sim}$20 of these systems are likely to be visible in the X-ray band; whether they are detectable with LISA depends crucially on their distance and orbital period (see Figure 9). Though we leave detailed predictions of the future evolution of the entire AIC population for future work, we note that the He-star donor AIC channel is an important formation channel for producing NS+WD mergers \citep[see also][]{toonen2018a}.

We remind the reader that our simulations assume field evolution only, e.g. no N-body interactions, no triple evolution etc. N-body interactions are most important in dense environments, such as globular clusters, thus the rate estimates presented in this work are most relevant for the Galactic disk.
Among higher stellar densities, stellar exchange interactions are likely to occur, which alters the evolutionary outcome of the stars. These additional interactions would likely lead to a higher number of AIC and MIC progenitor systems being created, which may be an important link toward understanding the formation of double NS mergers such as GW170817  \citep[see][figure 3]{belczynski2018a}. 

\section*{Acknowledgements}
AJR has been supported by the Australian Research Council through grant number FT170100243, and project number CE110001020 (Centre of 
Excellence for All-sky Astrophysics; CAASTRO). KB acknowledges support from the Polish National Science Center (NCN) grant:
Sonata Bis 2 DEC-2012/07/E/ST9/01360. 
IRS acknowledges funding from the 
Australian Research Council under grant FT160100028. 
AJR thanks Stuart Sim for helpful comments and Bernhard M\"{u}ller, Jarrod Hurley, Tyrone Woods, and Christopher Berry for discussion. Parts of this work made use of WebPlotDigitizer version 4 https://automeris.io/WebPlotDigitizer. LISA curves were constructed using the online sensitivity curve generator maintained by Shane Larson: http://www.srl.caltech.edu/$\sim$shane/sensitivity/.
This research was undertaken with the assistance of resources and services from the National Computational Infrastructure (NCI), which is supported by the Australian Government.


\section{Appendix} 

In this section we describe the evolutionary pathway leading to the formation of MIC and AIC progenitors in more detail. We focus on 4 specific examples. 

{\em Example 1: Classical CE model, AIC, medium-long delay time:}
The following evolutionary example showcases the evolution of an AIC system that evolves into a NS+HeWD binary, though with an orbital period that is larger than those typical for redbacks. 
The ZAMS masses are 7.2 and 1.4 \msun\, with a separation of 2310 \rsun. 
The first interaction occurs at 53 Myr after starburst when there is a CE phase between the late AGB primary (5.8 \msun) and the main sequence (1.4 \msun) companion. The CE phase causes the orbit to decrease from 
1923 \rsun\ to 31 \rsun\ 
and the AGB core evolves into a ONe WD. The next interaction doesn't occur until 3409 Myr after starburst when there is stable Roche-lobe overflow between the red giant (1.4 \msun) secondary and the ONe WD (1.3 \msun). Mass transfer lasts a few Myr, initially proceeding on a thermal timescale, until AIC occurs at 3411 Myr resulting in a NS (1.26 \msun) and the red giant is left with a mass of 0.96 \msun\ (with a core mass of 0.25 \msun) with an orbital separation of 38 \rsun\ (18 days).
A few Myr later, mass transfer is re-initiated when the red giant fills its Roche-lobe again. Mass transfer ceases when the stars have attained an orbital separation of 150 \rsun\ (156 days) and the masses are 1.5 \msun (NS) and 0.35 \msun\ (red giant; core mass 0.32 \msun). The secondary shortly thereafter evolves into a helium WD (0.34 \msun). The system is found at a Hubble time with component masses 1.5 \msun\ (NS), 0.34 \msun\ (He WD) with an orbital separation of 160 \rsun.  

{\em Example 2: Classical CE model, MIC, very long delay time:}
We describe an example of a system that attains very high gravitational wave frequencies (final orbital period ${\lesssim}$10s of seconds or $f_{\rm gw}>0.08$;  too rapidly-evolving to be found in Fig. 9) at time of the WD-WD merger. The chirp mass of this system falls near the second peak in Fig. 7, at $M_{\rm chirp}=1.06$. 
The ZAMS masses are 8.5 and 7.0 \msun\, with a separation of 62 \rsun. 
Stable Roche-lobe overflow starts between the Hertzsprung Gap primary and the MS companion at 33 Myr post-starburst. RLOF continues until the orbital separation reaches 340 \rsun\ at which time mass transfer ceases. By this time, the evolved primary is 1.7 \msun\ and the secondary has attained a mass of 10.3 \msun. At 40 Myr, mass transfer is re-established between the H-envelope stripped (via previous mass transfer), helium-burning sub-giant which fills its Roche-lobe and the MS secondary, and RLOF continues for a short time ($< 1$ Myr) until an orbital separation of 640 \rsun. The primary shortly thereafter evolves into a CO WD. At 46 Myr the secondary starts to evolve up the red giant branch, and at an orbital separation of 442 \rsun\ there is a CE between the red giant secondary and the CO WD. The system then consists of stars with masses 1.2 \msun\ (CO WD), 2.3 \msun\ (H-envelope stripped, He-burning star), and an orbital period of 3.9 \rsun. The secondary shortly thereafter becomes an evolved H-stripped, He-burning star and initiates RLOF toward the CO WD primary at 50 Myr. Mass transfer steadily continues until the secondary evolves into a ONe WD. The MIC progenitor is found with an orbital separation of 3.9 \rsun\ and component masses 1.2 and 1.2 \msun\ for the CO and ONe WD. The system loses orbital angular momentum due to gravitational wave emission and the stars merge at 10097 Myr. 
We note that this system falls above the maximum threshold assumed for NS masses in \citet{margalit2017a}, making it a viable stellar mass black hole candidate. 

{\em Example 3: New CE model, MIC, short delay time:} This system appears in the chirp mass distribution peak (see Fig. 8). 
In this short delay time channel example, the ZAMS masses are 7.4 and 5.2 \msun\, with a separation of $818$ \rsun. The initial primary fills its Roche lobe while in the Hertzsprung Gap and the main sequence secondary accretes matter stably, even after the primary evolves up the Red Giant Branch. After the orbit has become very wide (445 d) mass transfer ceases, and soon after the primary becomes a H-envelope-stripped, helium-burning star. The secondary, now more massive than its ZAMS mass, becomes a Red Giant and when it fills it Roche lobe, mass transfer is unstable, and a CE phase follows. Upon emerging from the CE, the system consists of a double helium-burning star binary where both stars have lost their envelopes, e.g. a double sdB star \citep[see][for observed examples of similar, lower-mass analogue systems]{kupfer2015a}, with an orbital period of 6.6 days and stellar masses of 1.27 and 1.91 \msun, respectively. Shortly thereafter the primary evolves into a (helium) giant, and transfers mass stably to the (still compact) helium-burning star companion. After the primary evolves into a CO WD, and the secondary becomes a (helium) giant, the system soon encounters a second CE phase where, yet again, the secondary star loses its envelope (this time its helium-rich envelope). Upon emerging from the CE, the binary consists of a CO WD and an ONe WD (masses 0.85 and 1.21 \msun, respectively) with an orbital period of 24 minutes. 

{\em Example 4: New CE model, AIC, short delay time:}
This example showcases a common channel in which the NS is formed via stable Roche-lobe overflow accretion on a ONe WD from the wind of an evolved AGB star. The stars start out on the ZAMS with masses 7.6 and 5.6 \msun\ with an orbital separation of 6150 \rsun. The primary evolves into a ONe WD at 47 Myr; by this time the orbit has increased in size to 10652 \rsun (stellar masses are 1.37 and 6.63 \msun). At 81 Myr post-starburst, the secondary becomes an evolved (thermally-pulsating) AGB star (6.4 \msun; core mass 1.2 \msun), and at this stage the ONe WD companion begins to accrete some mass lost in the AGB wind. By 82 Myr, the primary turns into a NS via AIC. The NS continues to accrete via the AGB wind, until the AGB star evolves into a CO WD (within 0.1 Myr). The final component masses are 1.28 and 1.23 \msun\ for the NS and CO WD, respectively, on an orbit with separation of 24 days.  

\newpage

\bibliographystyle{mn2e}
\bibliography{ashbibz.bib,amanda.bib,ashbibz-Lilia.bib}

\end{document}